\documentclass[12pt]{article}

\usepackage{amsmath}

\begin{document}

\input epsf

%
%

%
\newcommand{\Uno}{{\rm{}1\hspace{-0.9mm}l\hspace{-1.0mm}}}
\newcommand{\nat}{{\rm{}I\hspace{-0.4mm}l\hspace{-1.1mm}N}}
\newcommand{\entero}{\rule[2.2mm]{0.1mm}{0.5mm}
\hspace{-0.3mm}{\sf{}Z}\hspace{-1.7mm}
{\sf{}Z}\hspace{-0.3mm}\rule{0.1mm}{0.5mm}}
\newcommand{\complex}{\mbox{\rm\hspace{1.2mm}
\rule[0.05mm]{0.1mm}{2.5mm}\hspace{-1.2mm}C}}

\begin{titlepage}
\vspace{-1.5cm}
\hspace{11cm}MIT-CTP 3267
\vspace{1.cm}
\begin{center}

{\huge\bf{Gerbes and Duality}}\\
\vspace{1.0cm}
{\large\em{M. I. Caicedo}}\footnotemark[1]\footnotemark[2]
\vskip.2cm
Center for Theoretical Physics,\\
Laboratory for Nuclear Physics and Department of Physics,\\
Massachusetts Institute of Technology\\
Cambridge, Massachusetts 02139 USA\\

and\\

{\large\em{I. Mart\'{\i}n\footnotemark[3] and A.
Restuccia\footnotemark[4]}}
\vskip.2cm
Departamento de F\'{\i}sica\\
Universidad Sim\'{o}n Bol\'{\i}var.\\
Apartado postal 89000, Caracas 1080-A, Venezuela.
\end{center}

\footnotetext[1]{caicedo@lns.mit.edu also mcaicedo@fis.usb.ve}

\footnotetext[2]{On sabbatical leave from Departamento de
F\'{\i}sica, Universidad Sim\'on Bol\'{\i}var.}

\footnotetext[3]{isbeliam@usb.ve} \footnotetext[4]{arestu@usb.ve}

\end{titlepage}

\subsection*{\\Abstract}

\noindent{\small{
We describe a global approach to the study of duality
transformations between antisymmetric fields with transitions and
argue that the natural geometrical setting for the approach is
that of gerbes, these objects are mathematical constructions
generalizing U(1) bundles and are similarly classified by
quantized charges. We address the duality maps in terms of the
potentials rather than on their field strengths and show the
quantum equivalence between dual theories which in turn allows a
rigorous proof of a generalized Dirac quantization condition on
the couplings. Our approach needs the introduction of an auxiliary
form satisfying a global constraint which in the case of $1$-form
potentials coincides with the quantization of the magnetic flux.
We apply our global approach to refine the proof of the duality
equivalence between d=11 supermembrane and d=10 IIA Dirichlet
supermembrane.}}

\section{Introduction}



The usual electromagnetic duality concept first introduced by
Dirac in his dissertation on magnetic monopoles, later extended by
Montonen and Olive and used lately by Seiberg and Witten
\cite{Witten} to discuss the strong and weak coupling limits of
the low energy effective action of $N=2$ SUSY $SU(2)$ Yang-Mills
Theory, has provided a breakthrough in the understanding of the
non-perturbative analysis of QFT. It has also given a powerful
tool to unify different superstring and supermembrane theories and
to possibly  merge them in the context of M-theory, a theory of
membranes and $5$-branes whose low energy effective action is
$d=11$ supergravity {\cite{Ferrara}}. In Maxwell's theory
strong-weak coupling duality usually referred to as $T$-duality,
may be understood as a map between two quantum equivalent $U(1)$
gauge theories, one of them formulated in terms of a $U(1)$
$1$-form connection $A$ and coupling constant $\tau$ and its dual
theory given by another $U(1)$ $1$-form connection $V$ and
coupling constant $\frac{1}{\tau}$, the dual map being
intrinsically non-perturbative \cite{Witten2}.

Duality with $p$-forms with $p>1$ was first studied by Barb\'on in
\cite{Barbon} but his results only apply to globally defined
$p$-forms. In this article we want to extend those results
presenting the most general duality map between locally defined
$p$-forms, i.e. antisymmetric fields having non trivial
transitions on intersections of the open sets of a covering of a
compact $d$-dimensional manifold, in order to achieve this goal,
we must introduce the notion of $p$-gerbes. These are geometrical
objects which naturally describe the quantized charges associated
to antisymmetric fields and are consequently of interest for
$D$-brane and $p$-brane theories where the charges have a
topological origin.

Gerbes were first introduced by Giraud~\cite{Gi71} who was
studying non-abelian cohomology. Since then, they have been
carefully studied in the mathematical literature~\cite{Bry93}
\cite{Ch98} \cite{Gaj97} \cite{Hi99} unfortunately, only
abelian gerbes have been developed into a full geometrical theory
so far, while progress in the non abelian case although limited and
very recent, look quite promising for physical applications~\cite{Roman02} .

$p$-gerbes are geometrical structures that generalize $U(1)$
principal bundles with connection, in fact, they are the natural
setting to allow differential forms to have transitions in much
the same way that the $U(1)$ connection does. Gerbes allow the
consistent transitioning of $p+1$-order forms by promoting the
usual cocycle condition on the intersection of three open sets to
a $p+1$-cocycle condition on the intersection of $p+3$ open sets,
according to this convention, a line bundle is a $0$-gerbe. By
costruction, gerbes constitute a sort of ``geometrical ladder'' in
which a line-bundle ($0$-gerbe) is given by a set of transition
functions, a $1$-gerbe is given by a set of transition
line-bundles, a $2$-gerbe is given by a set of transition gerbes,
and so on ~\cite{Ch98} \cite{Hi99}. Gerbe-connections and
gerbe-curvatures can be defined by generalization of the
corresponding objects for bundles, the curvature of a $0$-gerbe
(i.e. of an ordinary connection) is a $2$-form, the curvature of a
$1$-gerbe is given by a $3$-form the definition obviously
extending up the geometrical ladder~\cite{portus}. These
gerbe-connections-curvatures share common issues with line-bundle
connections. The Kostant-Weil theorem for example has a gerbe
analogue~\cite{Bry93}\cite{Ch98}, while the first states that
line-bundles ($0$-gerbes) on ${\cal{M}}$ are classified by
$H^2({\cal{M}},\entero)$ (the second De Rham integer cohomology
over ${\cal{M}}$), the latter establishes that equivalence classes
of $1$-gerbes on ${\cal{M}}$ are classified by
$H^3({\cal{M}},{\entero})$, while in general $p$-gerbes are
classified by $H^{p+2}({\cal{M}},{\entero})$. $p$-gerbe connection
have an associated notion of parallel transport, the parallel
transport of an $p$-gerbe connection is defined along $p+1$
dimensional paths~\cite{Gaj97}\cite{Gaj99}\cite{portus}. This in
turn brings in the idea of holonomy, in the case ordinary
connections on line bundles holonomy associates a group element to
each loop while for the case of a $1$-gerbe-connection holonomy
associates a group element to each $2$-loop, some seminal work on 
this direction was presented in~\cite{NepoNPB}.
%

As we said before, the goal of this work is to study duality maps
between $p$-forms with transitions showing that gerbes provide a
natural setting for the problem. We shall address the duality in
terms not of the field strengths but on their potentials $A_p$ and
a $d-p-2$ form and will be able to rigourously obtain the
generalized Dirac quantization condition on the couplings
$g_pg_{d-p-2}=2\pi{}n$~\cite{NepoPRD}. To show the quantum equivalence between
dual electromagnetic theories, one starts from a theory defined
over the space of all connection $1$-forms on all line bundles
over the base $4$-manifold ${\cal{M}}$. The next step in the
process consists in building another theory with a Maxwell's like
action but in terms of globally constrained $2$-forms, at this
point another line bundle (${}^{*}L$) is introduced in the game to
allow the global contraints to be imposed via Lagrange
multipliers. After functional integration of the lagrange
multipliers represented by the connection form of ${}^{*}L$ the
dual theory is straightforwardly obtained. In the scheme just
described the global constraints are critical since they allow the
use of Weil's theorem guaranteeing the existence of a line bundle
with connection. The approach may be synthesized in the following
sequence:
\begin{eqnarray*}
A\Leftrightarrow\Omega_2(constrained)\Leftrightarrow{}V
\end{eqnarray*}
\noindent{}where $\Omega_2 $ is the globally defined constrained
2-form.

For $p$-forms over gerbes the approach to duality follows similar
lines, i.e. duality is shown through a similar sequence of steps
which we summarize in the following diagram
\begin{eqnarray*}
A_p \Leftrightarrow\Omega_{p+1}(constrained)\Leftrightarrow
V_{d-p-2}
\end{eqnarray*}
in this latter scheme, $A_p$ and $V_{d-p-2}$ represent the dual
antisymmetric fields, while $\Omega_{p+1}$ which is constrained to
be a closed form with integer periods plays an intermediate role
allowing the proof of duality. The constraint on the periods of
$\Omega_{p+1}$ is clearly global and consequently it must be
implemented ab initio in the mechanism to prove on shell global
equivalence and quantum equivalence between dual theories.

For dual maps between $1$-form connections in four dimensions the
global restriction coincides with the quantization of the magnetic
flux. In general, the global condition leads to important
relations between the relevant physical parameters involved in the
duality map. In the case of the $d=11$ supermembrane
$\Leftrightarrow$ $d=10$ Dirichlet supermembrane equivalence, the
global constraint becomes the compactification condition on one of
the supermembrane coordinates and we can use our machinery to
approach this problem.
%

The presentation of the work is as follows, section \ref{sec:u1}
motivates our general programme by carefully reviewing
electromagnetic duality emphasizing the role of the above
mentioned {\it{global}} constraint. In section
\ref{sec:introgerbs} we give a brief introduction to gerbes. In
section \ref{sec:HOB} we show that a antisymmetric fields with
quantized fluxes naturally give rise to gerbes, this property
being of importance for the implementation of our strategy to
duality. In section \ref{sec:HOBduality} we formulate the general
dual map between actions for $p$ and $p-d-2$ forms and show the
quantum equivalence of the dual theories. Finally, in section
\ref{sec:string} we apply our results to discuss duality maps in
$p$-brane theories.

\section{Duality on the space of connections on line bundles}
\label{sec:u1}

The purpose of this section is twofold, in the first place, we
will construct the quantum formulation of Maxwell's theory in
terms of a {\it{globally}} constrained $2$-form and explicitly
show its equivalence to the usual connection formulation. In
second place, using the above formulation we will show  the
duality between two $U(1)$ theories, one of them with coupling
constant $\tau$ and the other with coupling $-\frac{1}{\tau}$.

We begin by considering Maxwell's theory, formulated in terms of a
connection 1-form $A$ of a $U(1)$ bundle $L$ with base space
${\cal{M}}$ -a four dimensional compact orientable euclidean
manifold-, the theory is then given by the following action
\begin{equation}
I(F(A))=\frac{1}{8\pi}\int_{{\cal{M}}}d^4x\sqrt{g}[\frac{4\pi}{e^2}F^{mn}F_{mn}+
i\frac{\theta}{4\pi}\frac{1}{2}\epsilon_{mnpq}F^{mn}F^{pq}]
\end{equation}
\noindent{}where ($F=dA$) is the curvature associated to the
connection.

Duality is usually addressed in terms of the action of the modular
group $SL(2,\entero{})$ on the complex coupling constant
$\tau\equiv\frac{\theta}{2\pi}+i\frac{4\pi}{e^2}$. Upon
introduction of $\tau$ and using the standard decomposition of the
curvaturein its self-dual and anti self-dual parts, $I(F(A))$ can
be reexpressed as follows
\begin{equation}
I_\tau(A)=\frac{i}{8\pi}\int_{\cal{M}}d^4x\sqrt{g}[\bar{\tau}F_{mn}^+F^{+mn}
-\tau{}F_{mn}^-F^{-mn}]\ ,
\end{equation}
\noindent{}or in terms of the inner product of forms
\begin{equation}
I_\tau(A)=\frac{i}{4\pi}[\bar{\tau}(F^+,F^+)-\tau{}(F^-,F^-)]
\label{eq:itau}
\end{equation}
We now introduce an similar looking action but now the independent
field is a $2$-form $\Omega$ whose only property consisits in
being globally defined over ${\cal{M}}$, the action for $\Omega$
is then
\begin{equation}
{\cal{I}}(\Omega;\tau)=\frac{i}{4\pi}[\bar{\tau}(\Omega^+,\Omega^+)-\tau{}(\Omega^-,\Omega^-)]
\label{eq:iomeg}
\end{equation}
The quantum field theories associated to actions (\ref{eq:itau})
and (\ref{eq:iomeg}) are clearly inequivalent since $\Omega$ is
arbitrary, i.e. ${\cal{I}}(\Omega;\tau)$ is a functional over the
whole space of $2$-forms, while $F$ in $I_\tau(A)$ is the
curvature of a $U(1)$ connection.

In order to fulfill our programme of constructing quantum
Maxwell's theory in terms of a globally constrained $2$-form we
will show that after restricting the space of $2$-forms in
${\cal{I}}(\Omega;\tau)$ by the introduction of two constraints,
the theories defined by (\ref{eq:itau}) and the constrained
version of (\ref{eq:iomeg}) are equivalent as QFTs.

The constraints to be imposed on $\Omega$ are
\begin{eqnarray}
d\Omega{}=0\label{eq:locvinc}\\
\oint_{\Sigma_2}\Omega=2\pi{}n\label{eq:globvinc}
\end{eqnarray}
\noindent{}where $\Sigma_2$ represents a basis of the integer
homology of dimension 2 over ${\cal{M}}$. The first of these
constraints restricts $\Omega$ to be closed, while the second
ensures its periods to be integers (quantization of the ``magnetic
flux").

The first step in the discussion is to show that if one introduces
a new line bundle -to which we will refer to as  the dual line
bundle $^*\!{}L$- with connection $1$-form $V$, it is possible to
extend the action ${\cal{I}}(\Omega;\tau)$ in order to include
constraints (\ref{eq:locvinc}) and (\ref{eq:globvinc}) through the
appropriate use of $V$ as a Lagrange multiplier in the following
way
\begin{equation}
{\cal{I}}(\Omega,V;\tau)={\cal{I}}(\Omega;\tau)+
\frac{i}{2\pi}\int_{\cal{M}}W(V)\wedge\Omega
\label{eq:ivinc}
\end{equation}
\noindent{}where, $W(V)\equiv{}dV$ is the curvature associated to
$V$. Before we engage in the rigorous proof of the above claims,
we would like to note that using
$\int_{\cal{M}}W(V)\wedge\Omega=\int_{\cal{M}}dV\wedge\Omega$ as
the constraint in the extended action ${\cal{I}}(\Omega,V;\tau)$
instead of the usual formula $\int_{\cal{M}}V\wedge{}d\Omega$ is
critical, since as we will show, in the latter case, the
constraint in the periods of $\Omega$, which is a global
condition, would have never been obtained.

We begin by considering constraints (\ref{eq:locvinc}) and
(\ref{eq:globvinc}). If $F(A)$ is the curvature associated to a
connection $1$-form then it is obviously closed, i.e. satisfies
the local constraint (\ref{eq:locvinc}); moreover, the requirement
on the transition functions of the line bundle to be uniform maps
over the structure group guarantees that $F(A)$ also satisfies the
global constraint on its periods. The following
proposition~\cite{weil} which is a part of the Konstant-Weil
theorem, shows that the converse is also true:

{\it{If $\Omega$ is a $2$-form satisfying constraints
(\ref{eq:locvinc}) and (\ref{eq:globvinc}) then there exists a
complex line bundle and a connection -not necessarily unique- on
it whose curvature is $\Omega$}}

We will briefly review the proof of the above proposition since we
will closely follow it when dealing with higher order $p$-forms.
The proof is the following: let ${\cal{U}}=\{U_i\}$, $i\in{}I$
($I$ a set of indices) be a contractible open covering of
${\cal{M}}$. The condition of closeness on $\Omega$ guarantees
that it may be locally expressed as the exterior derivative of a
$1$-form, in particular, in a triple intersection of open sets
$U_i\bigcap{}U_j\bigcap{}U_k\neq\emptyset$ , $\Omega$ may be
written as:
\begin{equation}
\Omega=d{\cal{A}}_i=d{\cal{A}}_j=d{\cal{A}}_k
\end{equation}
\noindent{}where ${\cal{A}}_j$ represents an appropriate one form
locally defined in $U_j$, this in turn implies that the local
forms must be related by changes given by
\begin{eqnarray}
{\cal{A}}_i={\cal{A}}_j+d\Lambda_{ij}\\
{\cal{A}}_j={\cal{A}}_k+d\Lambda_{jk}\\
{\cal{A}}_k={\cal{A}}_i+d\Lambda_{ki}
\end{eqnarray}
\noindent{}where now the $\Lambda$'s are local $0$-forms. This
last set of identities lead to conclude that
\begin{equation}
\Lambda_{ij}+\Lambda_{jk}+\Lambda_{ki}=constant=c
\label{eq:suma-de-lambdas=entero}
\end{equation}
Finally, the global condition on the periods of $\Omega$ leads to
(see section \ref{sec:HOB} for details),
\begin{equation}
c=2\pi{}n
\end{equation}

The conclusion of these steps is clearly that in the sense of
\v{C}ech the cochain \cite{Steenrod}
\begin{equation}
g:(i,j)\rightarrow{}g_{ij}\equiv{}e^{i\Lambda_{ij}}\in{}U(1),
\end{equation}
\noindent{}is a $1$-cocycle
\begin{equation}
\delta{}g_{ijk}=g_{ij}g_{jk}g_{ki}=\Uno.
\end{equation}
Moreover, if the local $1$-forms are changed in intersecting open
sets by gauge transformations
\begin{eqnarray}
{\cal{A}}_i&\rightarrow&{\cal{A}}_i+d\lambda_i\quad\mbox{in}\quad{}U_i
\label{eq:2.14a}\\
{\cal{A}}_j&\rightarrow&{\cal{A}}_j+d\lambda_j\quad\mbox{in}\quad{}U_j
\label{eq:2.14b}
\end{eqnarray}
\noindent{}the glueing $0$-forms must change as
\begin{eqnarray}
\Lambda_{ij}\rightarrow{}\Lambda_{ij}+\lambda_i-\lambda_j
\label{eq:delta1}
\end{eqnarray}
\noindent{}implying that $g_{ij}$ changes as
\begin{eqnarray}
g_{ij}\rightarrow{}h_ig_{ij}h^{-1}_j.
\end{eqnarray}
One then notice that $h_ih_j^{-1}$ is a coboundary as follows from
the fact that
\begin{eqnarray}
h:(i)\rightarrow{}h_i=e^{i\lambda_i}\in{}U(1)
\end{eqnarray}
\noindent{}is a map from $U_i$ to the structure group, and
\begin{eqnarray}
\delta{}h(i,j)=h_ih_j^{-1}
\end{eqnarray}
\noindent{}consequently, under (\ref{eq:delta1}) $g_{ij}$ changes
by a coboundary, and then after the change it still defines the
same element of the \v{C}ech cohomolog
$\check{H}^1(\cal{U},\complex^*)$, where $\complex^{*}$ is the set
of non zero complex numbers. It is known \cite{Steenrod} that
there is a one-to-one correspondence between
$\check{H}^1(\cal{U},\complex^{*})$ and the complex line bundles
over ${\cal{M}}$, $g_{ij}$ defining the transition functions of
the bundle, therefore, constraints (\ref{eq:locvinc}) and
(\ref{eq:globvinc}) then define an unique line bundle over
${\cal{M}}$. Moreover ${\cal{A}}$, defined by patching together
the $1$-forms ${\cal{A}}_{i}$ by using
(\ref{eq:2.14a})(\ref{eq:2.14b}), is a connection $1$-form over
${\cal{M}}$ and $\Omega$ its curvature $2$-form.

Regarding the non uniqueness of the connections on the line bundle
associated to $\Omega$, one must realize that two connection
$1$-forms ${\cal{A}}^{(1)}$ and ${\cal{A}}^{(2)}$  with the same
curvature may be in different equivalence classes not related by
gauge transformations. They differ at most by a closed 1-form
$\theta\in{}H^1({\cal{M}},\Re)$. If $\theta$ is an element of
$H^1({\cal{M}},\entero)$ then ${\cal{A}}^{(1)}$ and
${\cal{A}}^{(2)}$ are connections on the same equivalence class
but otherwise they belong to different ones. The equivalence
classes of connections related to the same $\Omega$ are in
one-to-one correspondence  to $ H^1({\cal{M}},U(1))$. Moreover,
one has for the holonomy maps $Q$ constructed with connections
with the same curvature $\Omega$,
\begin{eqnarray*}
Q^{\chi{}.l}=\chi{}Q^l
\end{eqnarray*}
here $l$ denotes a line bundle with a particular equivalence class
of connections and $\chi$ is the holonomy map given by the
exponential of the integral of $\theta$  around a closed curve.
For a simple connected base manifold ${\cal{M}}$ the  line bundle
associated to $\Omega$ is unique {\cite{weil}}.

The observation just made is relevant to the proof of the quantum
equivalence of the theories defined by $I_\tau(A)$ and
${\cal{I}}(\Omega)$ restricted by the constraints we have been
studying. Indeed, when formulating the quantum correlation
functions for either theory, one must carefully define the
functional measure in order to account for the ``zero modes", that
is the space $H^1({\cal{M}},\Re)$.

It is worth noticing  that -up to the definition of the measure-,
the equivalence of the quantum theories rests on the non local
constraint on the periods of the $2$-form $\Omega$. Indeed, since
the local restriction $d\Omega{}=0$ is not sufficient to guarantee
the existence of a line bundle and a connection with curvature
$\Omega$ there is no local formulation of Maxwell's theory
($I_\tau(A)$) in terms of a globally defined closed $2$-form
$\Omega$. And therefore, the global constraint that associates a
set of integers $\{n\}$ (the winding numbers or topological
charges) to the elements of a basis of homology of dimension $2$
to $\Omega$ is a must.

In order to continue with the proof of the quantum equivalence, we
come to study the formulation of the off shell Lagrange problem
associated to action (\ref{eq:iomeg}), and constraints
(\ref{eq:locvinc}) and (\ref{eq:globvinc}). We will show that
action ${\cal{}I}(\Omega;\tau)$ constrained by both the local
($d\Omega=0$) and global ($\int\Omega=2n\pi$) conditions and the
extended action ${\cal{}I}(\Omega,V;\tau)$ are equivalent quantum
mechanically when summation over all line bundles is considered in
the functional integral.

We first consider the extra piece in ${\cal{I}}(\Omega{},V)$ i.e.
\begin{equation}
{\cal{S}}_{Lagrange}=\frac{i}{2\pi}\int_{\cal{M}}dV\wedge\Omega{}
\end{equation}
\noindent{}where we must recall that $V$ is a connection $1$-form
on the dual bundle $^{*}L$. ${\cal{S}}_{Lagrange}$ can be
rewritten as
\begin{equation}
{\cal{S}}_{Lagrange}=-\frac{i}{2\pi}\int_{\cal{M}}V \wedge
d\Omega{}+\frac{i}{2\pi}\int_{\cal{M}}d(V\wedge\Omega{})
\label{eq:extra}
\end{equation}

The functional integration on $V$ may be performed in two steps.
We first integrate on all connections over a given complex line
bundle and then over all complex line bundles. The second term on
(\ref{eq:extra}) depends only on the transition function of a
given complex line bundle, while the first depends also on the
space of connections over the line bundle. Integration associated
to the first step yields the following factor
\begin{equation}
\delta{}(d\Omega{})
\end{equation}
\noindent{}on the functional measure.

It is convenient to rewrite the second term in (\ref{eq:extra}) as
\begin{eqnarray}
\frac{i}{2\pi}\int_{\cal{M}}d(V\wedge\Omega{})
&=&\frac{i}{2\pi}\int_{\Sigma_{3}}(V_+-V_-)\wedge\Omega{}=\nonumber\\
&=&\frac{i}{2\pi}\int_{\Sigma_3}d\xi_{+-}\wedge\Omega{}=i n
\int_{\Sigma_{2}}\Omega
\label{eq:2.22}
\end{eqnarray}
\noindent{}where $\Sigma_3$ stand for $3$-dimensional surfaces
living in the intersection of open sets where the transition of
the connection $1$-form $V$ takes place,
\begin{eqnarray}
V_+-V_-&=&d\xi_{+-}\\
g_{+-}&=&e^{i\xi_{+-}}
\end{eqnarray}
\noindent{}$g_{+-}$ being the transition function and $\xi_{+-}$
is, in general, a multivalued function.

Recalling that the summation must be over all line bundles one
finds that formula (\ref{eq:2.22}) brings in the following factor
to the measure of the path integral
\begin{equation}
\sum_m \delta{}(\int_{\Sigma_{2}}\Omega{}-2\pi{}m)
\end{equation}

\noindent{}where $\Sigma_{2}$ denotes a basis of an integer
homology of dimension $2$. We thus conclude that the Lagrange
problem associated to the action (\ref{eq:iomeg}) constrained by
(\ref{eq:locvinc})(\ref{eq:globvinc}) is indeed given by the
extended action (\ref{eq:ivinc}).

We now turn to the discussion of the full partition function
associated to the extended action ${\cal{I}}(\Omega,V;\tau)$. The
path integral that we want to calculate is given by
\begin{equation}
{\cal{Z}}(\tau{})=\sum_{m}
\int{\cal{D}}\Omega{\cal{D}}V{\hbox{Vol}}_{ZM}{\hbox{det}}(d_2)
\frac{1}{{\hbox{Vol}}G}e^{-{\cal{I}}(\Omega,V;\tau)}
\end{equation}
where as we have just learned , $\sum_m$ stands for summation over
all line bundles. ${\hbox{Vol}}_{ZM}$ is the volume of the space
$H^1({\cal{M}},R)$, det($d_2$) is the determinant of the exterior
differential operator on $2$-forms and Vol$G$ is the volume of the
gauge group. After performing the integration on $V$ as described
in the previous paragraphs, we obtain
\begin{equation}
{\cal{Z}}(\tau{})=\sum_m
\int{\cal{D}}\Omega{\hbox{Vol}}_{ZM}{\hbox{det}}(d_2)\delta{}(d\Omega{})
\delta{}(\int_{\Sigma_{2}}\Omega{}-2\pi{}m)
e^{-{\cal{I}}(\Omega;\tau)}
\end{equation}

The measure may now be reexpressed in terms of an integration on
the space of connections $A$ over the line bundle $L$ in the
following way
\begin{equation}
{\cal{Z}}(\tau{})=\sum_m\int{\cal{D}}\Omega{\hbox{Vol}}_{ZM}\int{\cal{D}}A
\frac{1}{{\hbox{Vol}}G}
\frac{\delta{}(\Omega{}-F(A))}{{\hbox{Vol}}_{ZM}}
\delta{}(\int_{\Sigma_{2}}\Omega{}-2\pi{}m)
e^{-{\cal{I}}(\Omega;\tau)}
\end{equation}

The factor $1/{\hbox{Vol}}_{ZM}$ that comes from reexpressing
$\delta{}(d\Omega{})$ in terms of $\delta{}(\Omega{}-F(A))$
exactly cancels the volume originally appearing in the functional
measure. Further integration in $\Omega$  produces the final
result
\begin{equation}
{\cal{Z}}(\tau{})={\cal{N}}\int{\hat{\cal{D}}}A\frac{1}{{\hbox{Vol}}G}e^{-I_\tau{}(A)}
\label{eq:YESS}
\end{equation}

Where $\hat{\cal{D}}A$ denotes integration over the space of
connections on all line bundles over ${\cal{M}}$. Since
(\ref{eq:YESS}) is the partition function for the action
$I_\tau{}(A)$, we have been able to show the quantum equivalence
of the three formulations of Maxwell's theory thus finishing the
first part of our programme.

Finally, we would like to briefly discuss the duality
transformations in the functional integral associated to Maxwell's
theory. We start from the action
\begin{eqnarray}
{\cal{I}}(\Omega,V)=\frac{i}{4\pi}[\bar{\tau}(\Omega^+,\Omega^+)-
\tau{}(\Omega^-,\Omega^-)]\nonumber\\
+\frac{i}{2\pi}(W^+(V),\Omega^+)+\frac{i}{2\pi}(W^-(V),\Omega^-),
\end{eqnarray}
from where it is possible to perform the functional
integration on $\Omega^+$ and $\Omega^-$ to get the known result
\cite{Witten2}
\begin{equation}
{\cal{Z}}(\tau{})={\cal N}
\tau^{-\frac{1}{2}B^-_2}\bar{\tau}^{-\frac{1}{2}B^+_2}{\cal{Z}}(-\frac{1}{\tau})
\label{eq:dualz}
\end{equation}
\noindent{}where $B^+_k$ and $B^-_k$ are the dimensions of the
spaces of selfdual and antiselfdual $k$ forms, this last formula
can be reexpressed in terms of the Euler characteristic $\chi$ and
the Hirzebruch signature $\sigma$ as
\begin{equation}
[Im(\tau{})^{\frac{1}{2}(B_0-B_1)}{\cal{Z}}(\tau{})]= {\cal N}
\tau^{-\frac{1}{4}(\chi{}-\sigma{})}\bar{\tau}^{-\frac{1}{4}(\chi{}+\sigma{})}
[Im(-\frac{1}{\tau})^{\frac{1}{2}(B_0-B_1)}{\cal{Z}}(-\frac{1}{\tau})
\label{eq:dualz2}
\end{equation}
${\cal N}$ is a factor independent of $\tau$ that depends on the
topology of ${\cal{M}}$.

We have thus been able to implement the duality transformations in
a rigorous way by including the global constraint and the
associated measure factors in the functional integral of the
Maxwell action over a general base manifold ${\cal{M}}$.

\section{Introducing Gerbes}
\label{sec:introgerbs}

In most of the standard literature, antisymmetric tensors fields
are described in terms of  global $p$-form potentials defined over
a manifold ${\cal{M}}$, since these forms are global they don't
transition as usual connections do and therefore it is not clear
how they may adequately describe generalizations of magnetic
monopoles. In order to describe antisymmetric fields in the most
general way fields with nontrivial transitions over the base
manifold must be accounted for, their ''curvature'' being a
globally defined $p+1$-form. These transitioning field
configurations are the ones responsible for the appearance of
topological charges, this latter observation being essential in
the description of $p$-branes and $D$-branes from a quantum field
theory point of view. The natural setting for the description of
the above mentioned $p$-form potentials is in terms of $p$-gerbes
which we will try to describe in this section. Recalling that, as
we mentioned in the introduction, gerbes constitute a sort of
geometrical ladder (a $0$-gerbe for instance being nothing but a
line bundle $L$) we will try to introduce gerbes by describing the
lowest steps in such ladder.

The simplest way to define gerbes is by specifying the data which
is needed to reconstruct them. To build a $0$-gerbe all that is
needed is a contractible open covering ${\cal{U}}$ of ${\cal{M}}$
and a set of transition functions
$g_{ij}:U_i\cap{U}_j\rightarrow{}U(1)$ satisfying the usual rules:
$g_{ii}=1$, $g_{ij}=g_{ji}^{-1}$ and $g_{ij}g_{jk}g_{ki}=1$ for
any nonempty triple intersection, the last rule obviously being
the \v{C}ech condition for a $1$-cocycle.

The next objects in the hierarchy are $1$-gerbes. Their defining
data are~\cite{Bry93}\cite{Hi99} an open covering of ${\cal{M}}$,
and a set of maps
\begin{equation}
g_{ijk}:U_i\cap{}U_j\cap{}U_k\rightarrow{}S^1
\end{equation}
defined on each triple intersection satisfying
\begin{equation}
g_{ijk}=g_{jik}^{-1}=g_{ikj}^{-1}=g_{kji}^{-1}
\end{equation}
and a $2$-cocycle condition, in the intersection of four open
sets, namely
\begin{equation}
\delta{}g=g_{jk\ell}g_{ik\ell}^{-1}g_{ij\ell}g_{ijk}^{-1}=1
\quad{}on\quad{}
U_i\cap{}U_j\cap{}U_k\cap{}U_\ell
\end{equation}

Similar definitions apply for higher order $p$-gerbes, i.e. their
data are given by $S_1$ valued maps $g_{i_0i_1\dots{}i_p}$ defined
on the intersection of $p+2$-sets and which satisfy a
$p+1$-cocycle condition on the intersection of $p+3$ open sets.

Gerbes are sufficiently well behaved objects as to allow
differential geometry, gerbe-connections and gerbe-curvatures can
be defined by properly generalizing the corresponding objects for
bundles. The curvature of a $0$-gerbe is obviously the curvature
$2$-form of the bundle, the curvature of a $1$-gerbe is given by a
closed $3$-form and in general the curvature of a $p$-gerbe is a
$p+2$ closed form. These closed forms are the natural descendants
of a tower of differential forms (of orders $0$ to $p+1$) which, as
we shall see in this section, do naturally define the gerbes and
which are thus referred to as the gerbe connection. $p$-gerbe
connection have an associated notion of parallel transport, the
parallel transport of an $p$-gerbe connection is defined along
$p+1$ dimensional paths~\cite{Gaj97}\cite{Gaj99}\cite{portus} and
allow to define holonomies, which in the case ordinary connections
on $0$-gerbes associate an element of $U(1)$ to each loop, while
for the case of a $1$-gerbe-connection the association is form
$2$-loops to the group. In any case, these parallel transport
holonomy notios for $p$-gerbes call in the use of
categories\cite{portus} and are not of concern for this work.

One particularly interesting feature of gerbes is that in general
they are not manifolds, consequently we cannot point to them as
spaces as we do with line bundles. Fortunately, the notion of a
trivialization of a gerbe gives some insight about these
objects~\cite{Hi99}. In the case of a $0$-gerbe (a line bundle
$L$), a trivialization is a non-vanishing section $s$ of $L$. If
one choses $s$ to be unitary it is also a section of a principal
$S^1$-bundle, i.e. a collection of maps $f_i:U_i\rightarrow{}S^1$
where the open sets $U_i$ for a covering of the base space and
which in any intersection $U_i\cap{}U_j$ satisfies the rules
\begin{equation}
f_i=g_{ij}f_j
\end{equation}
where the $g_{ij}$ are the transition maps for the bundle. Given a
different trivialization $f^\prime_i$ the difference between the
two trivializations is defined as the family of quotients
$f^\prime_i/f_i$. By construction one finds that
\begin{equation}
\frac{f_i^\prime}{f_i}=\frac{f_j^\prime}{f_j}
\end{equation}
which shows that $f^\prime_i/f_i$ is the restriction to $U_i$ of a
global map thus showing that for a $0$-gerbe the difference
between two trivializations is nothing but a global map
$F:{\cal{M}}\rightarrow{}S^1$.

For $1$-gerbes a trivialization is defined by a set of functions
\begin{equation}
f_{ij}=f_{ji}^{-1}:U_i\cap{}U_j\rightarrow{}S^1
\end{equation}
such that $g_{ijk}=f_{ij}f_{jk}f_{ki}$. Once again, the difference
between two trivializations is given by the quotients
$h_{ij}=f_{ij}^\prime/f_{ij}$, which in turn satisfy
\begin{equation}
h_{ij}h_{jk}h_{ki}=1
\end{equation}
this is no other but \v{C}ech condition for a $1$-cocycle, meaning
that the difference between two trivializations of a $1$-gerbe is
a line bundle.

To make the definition of gerbes rigorously complete we should
mention that there must be some independence on the choice of the
trivializations. For $0$-gerbes this is reflected on the fact that
a line bundle is an element of $\check{H}^1(X,\Re{})$, i.e. an
equivalence class of $1$-\v{C}ech cocycles. Given the proper
definition of equivalence, $1$-gerbes are bijectively related to
$\check{H}^2({\cal{M}},\underline{U}_{\cal{M}}(1))$ \cite{Hi99}.
In the general case the generalization of the notion of
equivalence calls for the use of categories, for our current
purposes we won't go into such details but rather refer to the
mathematical literature~\cite{Bry93}\cite{Hi99}.

An alternative -much clearer and closer to the physicist- picture
of gerbes can be given in terms of multiplets of forms, and can be
conveniently introduced, risking being reiterative, by briefly
reviewing the field theory formulation of $1$-form connections or
stated in other words, the theory of line bundles formulated in
terms of their connections.

The isomorphism classes of line bundles with connections may be
represented by equivalence classes of doublets ($A,g$) where $A$
is a $1$-form connection over the base manifold ${\cal{M}}$ and
$g$ are the transition functions with values in $\complex^{*}$
-the nonzero complex numbers-. The connection (potential) is built
form a set of local one forms ($A_i$) which on double
intersections must be related by the transitions as usual, i.e.
$A_{i}-A_{j}=g_{ij}^{-1}dg_{ij}$. The transition maps must satisfy
the standard conditions $g_{ii}=I$, $g_{ij}=g_{ji}^{-1}$ and the
cocycle condition ($\delta {}g)_{ijk}=g_{ij}g_{jk}g_{ki}=I$ which
in terms of the exponential (or $log$) map is usually cast as
$
\Lambda _{ij}+\Lambda _{ij}+\Lambda _{ij}=2\pi n,\mbox{ }n\in Z
$

The equivalence relation between to doublets is defined by stating
that two doublets are equivalent or gauge equivalent in physicists
language ($(A^{\prime},g^{\prime})\sim{}(A,g)$) if and and only
if, for any $U_i$ there is a map $h_i:U_i\rightarrow\complex^{*}$
such that the potentials and the transition functions are related
in the usual way as $A_{i}^{\prime }=A_{i}+h_{i}^{-1}dh_{i}$ and
$g_{ij}^{\prime }=h_{i}g_{ij}h_{j}^{-1}$ on $U_{i}\cap {}U_{j}$,
this relations are much more familiar if we express them in terms
of the arguments of the exponential map, which formulates the
transitions as in formulas (\ref{eq:2.14a}), (\ref{eq:2.14b}) and
(\ref{eq:delta1}). In such representation, the equivalence between
the transition $0$-forms shows that the integer appearing in
formula (\ref{eq:suma-de-lambdas=entero}) classifies the doublet,
the integer being the quantized flux
$\int{}F/2\pi\in{}H^2({\cal{M}},\entero)$. This is just the fact
that line bundles are classified by second integer De Rham
cohomology over ${\cal{M}}$ ($H^2({\cal{M}},\entero)$) which comes
from the isomorphism
$\check{H}^1({\cal{M}},\underline{U}_{{\cal{M}}}(1))\cong{}H^2({\cal{M}},\entero)$
which in turn follows from the exact sequence of sheaves
\begin{equation}
0\longrightarrow\entero_M\stackrel{\times{}2\pi{}i}
{\longrightarrow}\underline{\complex}_M\stackrel{\exp}{\longrightarrow}
\underline{\complex}^{*}_M\longrightarrow{}1
\end{equation}
and the isomorphism
$\check{H}^2({\cal{M}},\entero_{{\cal{M}}}(1))\cong{}H^2({\cal{M}},\entero)$
~\cite{Bry93}\cite{portus}.

Let us now describe the generalization of the doublet structure to
a set of triplets $(B,\eta{},\Lambda{})$, where $B$ represents a
$2$-form locally defined on the open sets of a covering, $\eta$ is
a local $1$-form defined on double intersections, and $\Lambda$ is
a $0$-form -the transition function for the $1$-forms- locally
defined on triple intersections (The generalization to $p$-plets
goes along the same lines).

Let $B_{i}$ stand for the local $2$-form as described in $U_i$, an
element of an open covering of ${\cal{M}}$, in analogy with the
connection one forms these $2$-forms must transition on the
intersection of two open sets, the only difference is that now the
transitions are given by the $1$-forms of the multiplet as
\begin{equation}
B_{i}-B_{j}=d\eta _{ij}
\label{eq:triplet1}
\end{equation}
the $1$-forms $\eta _{ij}$ in turn must be matched (must
transition) on the intersection of three open sets i.e. when
$U_{i}\cap {}U_{j}\cap {}U_{k}\neq\emptyset$
\begin{equation}
\eta_{ij}+\eta_{jk}+\eta_{ki}=d\Lambda_{ijk}  \label{eq:triplet2}
\end{equation}
while on $U_{i}\cap {}U_{j}\cap {}U_{k}\cap {}U_{l}\neq\emptyset$,
the $0$ forms $\Lambda $ satisfy the $2$-cocycle condition
\begin{equation}
(\delta \Lambda {})_{ijkl}=\Lambda _{ijk}-\Lambda _{ijl}+\Lambda
_{ikl}-\Lambda _{jkl}=2\pi {}n.
\label{eq:triplet3}
\end{equation}
(\ref{eq:triplet1}), (\ref{eq:triplet2}) and (\ref{eq:triplet3})
define the transition properties of a triplet $(B,\eta{},\Lambda)$
on a covering of ${\cal{M}}$. To give a complete description of
the triplet defining a $1$-gerbe we need to introduce a notion of
the equivalence between triplets.

Two triplets $(B,\eta {},\Lambda {})$ and $(\tilde{B},\tilde{\eta
},\tilde{\Lambda })$ if for each open set $U_i$ and each nonempty
intersection $U_i\cap{}U_j\neq\emptyset$ there exists $1$-forms
$\omega_i$ and $0$-forms $\Theta_{ij}$ such that the local forms
in each triplet are related by
\begin{eqnarray}
\hat{B}_i&=&B_i+d\omega_i\\
\label{eq:gaugetrip1}
\hat{\eta}_{ij}&=&\eta_{ij}+\omega_i-\omega_j+d\theta_{ij}\\
\label{eq:gaugetrip2}
\hat{\Lambda}_{ijk}&=&\Lambda _{ijk}+\theta _{ij}+
\theta_{jk}+\theta _{ki}
\label{eq:gaugetrip3}
\end{eqnarray}
rules (\ref{eq:gaugetrip1}), (\ref{eq:gaugetrip2}) and
(\ref{eq:gaugetrip3}) define the gauge transformations on the
space of triplets and clearly generalize those of the connection
and transition functions for a line bundle ($0$-gerbe). With these
definitions the equivalence classes of triplets are independent of
the covering, and the isomorphism
$\check{H}^{1}({\cal{M}},\underline{C}_{{\cal{M}}}^{*})\cong
{}\check{H}^{2}({\cal{M}},\entero)$ generalizes to the following
isomorphism between cohomology groups

\begin{equation}
\check{H}^{2}({\cal{M}},\underline{C}_{{\cal{M}}}^{*})
\cong\check{H}^{3}({\cal{M}},\entero)
\end{equation}
where the second \v {C}ech cohomology group
$\check{H}^{2}({\cal{M}},\underline{C}_{{\cal{M}}}^{*})$ is to be
identified with the $0$-form transitions satisfying the
$2$-cocycle condition, and $\check{H}^{3}({\cal{M}},\entero)$ is
related but not exactly isomorphic to the $3$-forms with integer
periods, and in fact it generalizes the Chern classes of doublets
to triplets.

The identification between gerbes and multiplets of fields was
first suggested by Deligne~\cite{Bry93}\cite{portus}, in the case
of $0$-gerbes, the equivalence classes of line bundles with
connection are in one-to-one correspondence to the cohomology
classes in the first smooth Deligne hypercohomology group:
\begin{equation}
{H}^{1}({\cal{M}},\underline{\complex}_{{\cal{M}}}^{*}
\stackrel{dlog}
{\longrightarrow}\underline{\cal{A}}^1_{\cal{M},\complex})
\end{equation}
while for triplets ($1$-gerbes with connection) the bijective
correspondence is to the second cohomology class of Deligne's
hypercohomology
\begin{equation}
{H}^{2}({\cal{M}},\underline{\complex}_{{\cal{M}}}^{*}
\stackrel{dlog}
{\longrightarrow}\underline{\cal{A}}^1_{\cal{M},\complex}
\stackrel{d}
{\longrightarrow}\underline{\cal{A}}^2_{\cal{M},\complex})
\end{equation}

These and other relevant exact sequences have been extensively
studied in~\cite{Hi99} and the important conclusion is that in
general $p$-gerbes are classified by
$H^{p+2}({\cal{M}},{\entero})$~\cite{portus}.

\section{Flux quantization and Gerbes}
\label{sec:HOB}

The purpose of this section is to show how a closed form with a
quantized flux naturally describes a gerbe in much the same way
that a magnetic monopoles describes a bundle with connection. In a
sense, we will be generalizing the reasoning behind Weil's
theorem. Before entering the subject we will introduce a unified notation,
a one indexed object (such as $A_{2}$) stands for a $2$ form, while
something like $\Lambda_{1i}$ is a $1$-form locally defined on an element
$U_i$ of a covering.

We consider closed integer forms globally defined over ${\cal{M}}$
-an orientable compact euclidean manifold- and explicitly show
that such forms have a natural geometrical structure associated them. This
geometrical structure consists of equivalence classes of
multiplets of local forms generalizing the local connection
$1$-forms and their related transition functions. For a
$p+1$-form $F_{p+1}$, the multiplet contains local $p$-forms with
values in the $U(1)$ algebra, and a tower of forms of decreasing order which
ends up in zero forms satisfying a $p$-cocycle condition in the
intersection of $p+2$ open neighborhoods of a covering of
${\cal{M}}$ and thus the strucutre is that of a $p-2$-gerbe. In
the particular case $p=3$ for example, a closed $3$ form $F_3$
with integer flux gives rise to triplets
$(A_{2i},\Lambda_{1i},\Lambda_{0i})$ of local forms defined in
each open set ${\cal{U}}_i$, i.e. a $1$-gerbe.

For the sake of completeness, we start our presentation with the
simplest case: $p=1$, this is relevant to show the equivalence
between the $d=11$ supermembrane and the $d=10$ IIA Dirichlet
supermembrane which we will adress in the next section.

To begin our discussion, let $F_1$ be a 1-form globally defined
over ${\cal{M}}$ satisfying
\begin{eqnarray}
dF_{1}&=&0\label{eq:l1closed}\\
\oint_{\Sigma_{1}}F_1&=&2\pi{}n,\quad{}n\in{\entero}\label{eq:l1isinteger}
\end{eqnarray}
where $\Sigma_1$ is a basis of an integer homology of curves over
${\cal{M}}$ and $n$ is an integer associated to each element of
the basis, then $F_1$ must be given by
\begin{equation}
F_{1}=-ig^{-1}dg\label{eq:3.3}
\end{equation}
where
\begin{equation}
g=\exp{i\varphi}\label{eq:3.4}
\end{equation}
defines an uniform map
\begin{equation}
{\cal{M}}\rightarrow S^{1},\label{eq:3.5}
\end{equation}
$\varphi$ being an angular coordinate on $S^1$.

Conversely, given $g$ an uniform map from ${\cal{M}}\rightarrow
S^1$ then the formula $-ig^{-1}dg$ does indeed define a closed
$1$-form with integer periods. To prove this claim, we begin by
realizing that with $F_1$ given as in (\ref{eq:3.3}), one may
define
\begin{equation}
\varphi(P)=\int_{O}^{P}F_{1}
\label{eq:phi}
\end{equation}
where $O$ and $P$ are the two end points of a curve on the base
manifold ${\cal{M}}$, $O$ being a reference point. $\varphi(P)$ is
independent of the curve within a homology class in the sense that
if ${\cal{C}}$ and ${\cal{C}^\prime}$ are open curves with the
same end points
\begin{eqnarray}
\int_{{\cal C}}F_{1}=\int_{{\cal C }^\prime}F_{1}
\end{eqnarray}
\noindent{}
if the closed curve ${\cal{C}}^{-1}{\cal{C}}^{\prime}$ is
homologous to zero and, by assumption \ref{eq:l1isinteger},
differs in $2\pi n$ between two different homology classes.
(\ref{eq:3.4}) then defines an uniform map from
${\cal{M}}\rightarrow{}S^1$. The converse follow directly by the
same arguments.

We may also understand the origin of $\varphi$ as given in
(\ref{eq:phi}) from a different point of view by considering a
covering of ${\cal{M}}$ with open sets $U_i$, $i\in{}I$. We may
always assume $U_i$ and $U_i\cap{}U_j\neq\emptyset$, $i,j\in{}I$
to be contractible to a point.

Since $F_1$ is closed it is locally exact, and thus on $U_i$,
$i\in{}I$
\begin{equation}
F_{1}=d\lambda_{0i}\quad{}\lambda_{0i}\mbox{ being a local
$0$-form},
\end{equation}
\noindent{}and on $U_i\cap{}U_j\neq\emptyset$
\begin{eqnarray}
d\lambda_{0i}&=&d\lambda_{0j}\\
\lambda_{0i}&=&\lambda_{0j}+c_{ji}
\end{eqnarray}
where $c_{ji}$ is a constant on the intersection.

On $U_j$ we may define
\begin{equation}
\lambda^{\prime}_{0j}=\lambda_{0j}+c_{ji}
\end{equation}
\noindent{}without changing $F_1$ and with a trivial transition on
$U_i\cap{}U_j\neq\emptyset$.

We may try continue this process of redefining $\lambda_{0i}$
through trivial transitioning in order to systematically extend it
to all the open sets on the covering, but at some point $U_j$, the
global condition on the periods of $F_1$ will impose an
obstruction to the process, and then we will only be able to write

\begin{equation}
F_{1}=d\lambda
\end{equation}
\noindent{}at the expense of dealing with a multivalued function
$\lambda$. Condition (\ref{eq:l1isinteger}) thus ensures that the
transition (which in this case defines the multivaluedness of
$\lambda$) is $2\pi{}n$, and we do consequently obtain
(\ref{eq:3.3}), (\ref{eq:3.4}) and (\ref{eq:3.5}).

%
%
Let us consider the degree $2$ case already discussed in section
\ref{sec:u1} to which we would like to add some remarks. Let $F_2$
be a $2$-form globally defined on ${\cal{M}}$ and let $F_2$ be
closed and satisfy the constraint on its periods (
$\oint_{\Sigma_{2}}F_{2}=2\pi{}n$) on a basis of an integer
homology of dimension $2$ ($\Sigma_2$). Then according to the
Konstant-Weil theorem there exists a complex line bundle with
conection with base space ${\cal{M}}$ whose curvature is $F_2$ and
conversely, given a line bundle over ${\cal{M}}$ with local
connection $1$-form $A_{1i}$ its curvature $F_2=dA_{1i}$ satisfies
is both closed and integer.

\begin{figure}[h]
\epsfxsize=2in \centerline{\epsffile{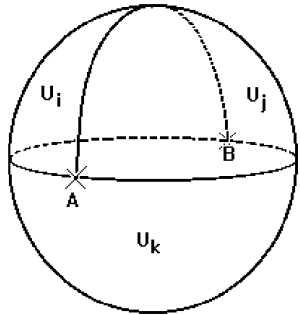}}
\caption{Three intersecting open sets}
\label{fig:interseccion}
\end{figure}

We are interested in the first statement, cuantized flux implies
the existence of a $0$-gerbe. As we showed in section \ref{sec:u1}
iterative use of Poincar\`e's lemma shows that in each open set of
a covering of ${\cal{M}}$ $F_2$ is given by a set of local
$1$-form $A_{1i}$, which on nonempty intersections of two elements
of the covering must be related by transition $0$-forms in the
usual fashion $A_{1i}=A_{1j}+d\Lambda_{0ij}$ the transition
$0$-forms in turn being forced to satisfy the condition:
$\Lambda_{0ij}+\Lambda_{0jk}+\Lambda_{0ki}=\mbox{constant}$ on
triple not empty intersections
($U_i\cap{}U_j\cap{}U_k\neq\emptyset$). We will now complete a
detail that we left open in section \ref{sec:u1}, the proof that
the constant in the latter condition is given by the period of
$F_2$, indeed, if we take a ``surface" $\Sigma_2$ in the
intersection $U_i$, $U_j$ and $U_k$, (see Figure
\ref{fig:interseccion}), we can calculate the period of $F_2$ to
get
\begin{equation}
\int_{\Sigma_2}F_{2}=\int_{\Sigma_{1}}A_{1}=
\displaystyle\left.(\Lambda_{0ij}+\Lambda_{0jk}+\Lambda_{0ki})
\right|_{B}^{A}{}=2\pi{}n
\end{equation}
\noindent{}where $\Sigma_1$ is the union of the three curves on
the figure, and
$A_1=d(\Lambda_{0ij}+\Lambda_{0jk}+\Lambda_{0ki})$. Without
loosing generality we may redefine the $\Lambda$'s in such a way
as to make the value of
$\Lambda_{0ij}+\Lambda_{0jk}+\Lambda_{0ki}$ at $B$ equal to zero,
and thus obtain the cocycle condition
\begin{equation}
\Lambda_{0ij}+\Lambda_{0jk}+\Lambda_{0ki}=2\pi{}n.
\label{eq:la-suma-es-entera}
\end{equation}

To complete the construction of the $0$-gerbe associated to $F_2$
equivalence classes of doublets of forms $(A_1,\Lambda_0)$ must be
adequately defined. As we already know, this is done by
introducing local $0$-forms $\lambda_{0i}$ and declaring
$(A_1,\Lambda_0)\sim(\tilde{A}_1,\tilde{\Lambda}_0)$ if
$\tilde{\Lambda}_{0ij}-\Lambda_{0ij}=\lambda_{0i}-\lambda_{0j}$
and $\tilde{A}_{1i}-A_{1i}=d\lambda_{0i}$

%
%
In what follows we will try to follow the steps we have just
taken, i.e. an iterative procedure involving Poincar\`es's lemma,
in order to build similar multiplets of forms for higher order
closed and globally defined forms. With this in mind, let us now
consider a $3$-form $F_3$ globally defined on ${\cal{M}}$
satisfying the constraints we have been considering, i.e.
closedness and the ``quantization" of its flux,
\begin{equation}
dF_{3}=0\qquad\mbox{and}\quad{}
\oint_{\Sigma_{3}}F_3=2\pi{}n,\quad{}n\in\entero\label{eq:F3-is-integer}
\end{equation}
where $\Sigma_{3}$ is a basis of an integer homology of dimension
$3$.

Since $F_3$ is closed we can build a local $2$ form $A_{2i}$ in
any subset $U_i$ of the covering in such a way as to have
\begin{equation}
F_{3}=dA_{2i}
\end{equation}
and since in the intersection of two sets
($U_i\cap{}U_j\neq\emptyset$)
\begin{equation}
dA_{2i}=dA_{2j}
\end{equation}
the local $2$ forms must by related (transition) as
\begin{equation}
A_{2i}=A_{2j}+ d\Lambda_{1ij}
\label{eq:transition_para_A2}
\end{equation}
where this time, the transition is given by, a $1$-form
$\Lambda_{1ij}$ which is locally defined on $U_i\cap U_j$.
Clearly, on a nonempty triple intersection $U_i\cap U_j\cap U_k$
it happens that
\begin{equation}
d(\Lambda_{1ij}+\Lambda_{1jk}+\Lambda_{1ki})=0
\end{equation}
meaning that in such triple intersection the sum of the transition
one forms is locally a $0$-form, i.e.
\begin{equation}
\Lambda_{1ij}+\Lambda_{1jk}+\Lambda_{1ki}=d\Lambda_{0ijk}
\end{equation}
If we now follow the same reasoning that was used to determine the
value $2\pi{n}$ for the cocycle condition in the study of $F_2$,
we can find the periods of $\Lambda_{1ijk}\equiv{}d\Lambda_{0ijk}$
on any closed curve on the triple intersection where it has been
defined. Considering an element $\Sigma_3$ of the $3$-homology of
${\cal{M}}$ intersecting $U_i$, $U_j$ and $U_k$, and using formula
(\ref{eq:F3-is-integer}) which states the condition on the periods
of $F_3$ we obtain
\begin{equation}
\int_{\Sigma_{3}}F_{3}=\int_{\Sigma_{1}}\Lambda_{1ijk}=2\pi n,
\label{eq:flujo-de-f3-f1-entero}
\end{equation}
where $\Sigma_{1}$ is a closed curve on $U_i\cap{}U_j\cap{}U_k$.
By construction $\Lambda_{1ijk}$ is both closed and with integer
periods in any closed curve in $U_i\cap{}U_j\cap{}U_k$ and
therefore defines a uniform map $M$
\begin{equation}
M:U_i\cap{}U_j\cap{}U_k \rightarrow{}U(1)\label{eq:3.18}
\end{equation}

At this point we find a new and interesting twist to the story.
Since the transitions of the $2$-forms are given by local one
forms on the intersections of the elements of the covering, we can
naturally build the following maps
\begin{equation}
g:(i,j)\rightarrow{}g_{ij}(P,{\cal{C}})\equiv
\exp{i\int_{\cal{C}}\Lambda_{1ij}}\label{eq:new-1-cochain}
\end{equation}
where ${\cal{C}}$ is an open curve with end points $O$ (a
reference point) and $P$. $g$ associates to $(i,j)$ a map
$g_{ij}(P,\cal C)$ from the path space over $U_i\cap U_j$ to the
group $U(1)$ and consequently, $g$ is a $1$-cochain.

Notice that the $1$-form $\eta_{ij}$ cannot be integrated out to
obtain a transition function as in the case of a line bundle.
However, if we apply \v{C}ech's coboundary operator to $g$ we
obtain
\begin{equation}
\delta{}g_{ijk}=g_{ij}g_{jk}g_{ki}=
\exp{}i\int_{O}^{P}\Lambda_{1ijk}\label{eq:not-a-cocycle}
\end{equation}
which is in general different form the identity element of $U(1)$,
and is in fact, the uniform map $M$ previously defined in
(\ref{eq:3.18}). The fact that the coboundary
(\ref{eq:not-a-cocycle}) is not the identity map on the group
explicitly shows that the geometrical structure we are dealing
with is not that of a $U(1)$ bundle. Nevertheless,
(\ref{eq:not-a-cocycle}) is a properly defined $2$-cochain in
\v{C}ech's cohomology theory. With this objectat hand, we may go a
step furhter and consider the action of the coboundary operator
$\delta$ on 2-cochains, and consequently we need to study what
happens in the intersection of four open sets. We will show that a
$2$-cocycle can be built in such intersection. Indeed, in the
intersection of four open sets ($U_i\cap U_j\cap U_k\cap U_l$)
\begin{eqnarray}
d\Lambda_{0ijk}&=&\Lambda_{1ij}+\Lambda_{1jk}+\Lambda_{1ki}\\
d\Lambda_{0ijl}&=&\Lambda_{1ij}+\Lambda_{1jl}+\Lambda_{1li}\\
d\Lambda_{0ikl}&=&\Lambda_{1ik}+\Lambda_{1kl}+\Lambda_{1li}\\
d\Lambda_{0jkl}&=&\Lambda_{1jk}+\Lambda_{1kl}+\Lambda_{1lj}
\end{eqnarray}
from where it follows that
\begin{equation}
d(\Lambda_{0ijk}-\Lambda_{0ijl}+\Lambda_{0ikl}-\Lambda_{0jkl})=0
\end{equation}
or equivalently,
\begin{equation}
\Lambda_{0ijk}-\Lambda_{0ijl}+\Lambda_{0ikl}-\Lambda_{0jkl}=\mbox{constant,}
\end{equation}
Using identity (\ref{eq:flujo-de-f3-f1-entero}) we finally obtain,
\begin{equation}
\Lambda_{0ijk}-\Lambda_{0ijl}+\Lambda_{0ikl}-\Lambda_{0jkl}=2\pi{}n
\end{equation}
It is then natural to define the following $2$-cochain on
$U_i\cap{}U_j\cap U_k$
\begin{equation}
g:(i,j,k)\rightarrow{}g_{ijk}\equiv\exp{i\Lambda_{0ijk}}
\end{equation}
which does obviously satisfy the $2$-cocycle condition
\begin{equation}
\delta{}g_{ijkl}=g_{ijk}g^{-1}_{ijl}g_{ikl}g^{-1}_{jkl}=I
\end{equation}

We thus conclude that a $3$-form with integer periods has an
associated triplet $(A_2,\Lambda_{1},\Lambda_{0})$ of local forms,
the latter of which satisfies a $3$-cycle condition on the
intersection of four open sets of the covering of the manifold,
i.e. a $1$-gerbe.

%
The procedure we have shown may be generalized to globally defined
$p+1$-forms $F_{p+1}$ over ${\cal{M}}$, satisfying
\begin{eqnarray}
dF_{p+1}&=&0\\
\oint_{\Sigma_{p+1}}F_{p+1} &=& 2\pi n.
\end{eqnarray}
This gives a geometrical structure with transition $p-1$ forms
$\Lambda_{p-1}$ with values on the Lie algebra of the structure group
leading to $1$-cochains

\begin{equation}
\exp{i\int_{\Sigma_{p-1}}\Lambda_{p-1}}
\end{equation}
\noindent{}$F_{p+1}$ being the field strenght (curvature) of a local $p$
form $A_{p}$ with transitions given by $d\Lambda_{p-1}$. Moreover on
$U_i\cap U_j\cap U_k\neq\emptyset$ the $p-1$-transition form
\begin{equation}
F_{p-2}=\Lambda_{p-1ij}+\Lambda_{p-1jk}+\lambda_{p-1ki}
\end{equation}
satisfies the conditions
\begin{eqnarray}
dF_{p-2} & =& 0\\
\oint_{\Sigma_{p-2}}F_{p-2} &=& 2\pi n
\end{eqnarray}
and hence the structure of $F_{p-2}$ may be determined by
induction. We end up with a $p$-cocycle condition on the
intersection of $p+2$ open sets. Summing up, we have shown that
the globally constrained $p+1$ form $F_{p+1}$ implies the existence
existence of a tower (multiplet) of local antisymmetric fields with
non trivial transition which in turn represents a $p$ gerbe.

In section~\ref{sec:string}, we will apply these lessons 
to show quantum
equivalence of the $d=11$ supermembrane and the $d=10$ IIA
Dirichlet supermembrane for the general case of non trivial line
bundles associated to the $U(1)$ gauge fields in the Dirichlet
supermembrane multiplet.  We will thus extend previous proofs
valid for trivial line bundles.

A last word about gerbes, as we stated in the introduction, 
gerbes only abelian gerbes have been developed into a full geometrical 
theory so far, while not much progress has been made in the non abelian
case. The reasons for this do probably date back to the work 
of Teitelboim~\cite{teit} who showed that non abelian theories 
for higher order forms do not exist. As we have seen in this section, the 
construction of gerbes in terms of multiplets of forms relies on a 
reasonable extension of Weil's theorem, a similar approach has been introduced 
in~\cite{isbe1}\cite{isbe2}, to discus the construction
of duality maps for non-abelian $2$-forms but such work is still in progress.

\section{Duality on $p$-Gerbes}
\label{sec:HOBduality}

In this section, we discuss the general duality map relating local
antisymmetric fields defining gerbes. The action for the local
$U(1)$ $p$-form $A_p$ defined over open sets of a covering of
${\cal{M}}$, a compact manifold of dimension $d\ge{}p+1$ with
$p>1$, and with transitions given as in the previous section, is
the following
\begin{equation}
S(A_p)=\frac{1}{2}\int_{\cal{M}}F_{p+1}\wedge{}^{*}F_{p+1}{}+
g_p\int_{\Sigma_{p}}A_p
\label{eq:s-p-form}
\end{equation}
where $F_{p+1}$ is the globally defined curvature $p+1$-form
associated to $A_p$. $\Sigma_p$ is a $p$-dimensional closed
surface being the boundary of a $p+1$-chain. $g_p$ is the coupling
associated to $A_p$. From (\ref{eq:s-p-form}) we obtain the field
equations
\begin{equation}
d^{*}F_{p+1}=g_p\delta_{\Sigma_p}
\label{eq:4.2}
\end{equation}
where $\delta_{\Sigma_p}$ is the usual $d-p$-form associated to
the Dirac density distribution.

Let us consider now the dual formulation to (\ref{eq:s-p-form}).
Following the arguments of the previous
sections, we introduce a constrained $p+1$-form $\Omega_{p+1}$
globally defined over ${\cal{M}}$ and satisfying
\begin{eqnarray}
d\Omega_{p+1}&=&0\\
\oint_{\Sigma_{p}}\Omega_{p+1}&=&\frac{2\pi{}n}{g_p}\label{eq:global-omega}.
\end{eqnarray}
we also introduce the following action for $\Omega_{p+1}$
\begin{equation}
I(\Omega_{p+1})\equiv{}
\frac{1}{2}\int_{\cal{M}}\Omega_{p+1}\wedge{}*\Omega_{p+1}{}+
g_p\int_{\Sigma_{p+1}}\Omega_{p+1}
\end{equation}
\noindent{}where $\Sigma_{p+1}$ is a $p+1$-chain with boundary
$\Sigma_{p}$.

The off-shell Lagrange problem  of the above constrained system
may be given by the extended action
\begin{equation}
{\cal{I}}(\Omega_{p+1},V_{d-p-2})=S(\Omega_{p+1})+
i\int_{\cal{M}}\Omega_{p+1}\wedge{}W_{d-p-1}(V){}
\end{equation}
where $W_{d-p-1}\equiv{}dV_{d-p-2}$ is the field strenght of the local
$d-p-2$-form $V_{d-p-2}$, i.e. the curvature of $d-p-2$-gerbe.
Consequently, $W_{d-p-1}$ identically satisfies the conditions
\begin{eqnarray}
dW_{d-p-1} & = & 0\\
\oint {W_{d-p-1}} &=& \frac{2\pi n}{g_{d-p-2}}.
\end{eqnarray}
Integration on $V_{d-p-2}$ leads to action
(\ref{eq:s-p-form}) while, integration on $\Omega_{p+1}$ yields
the on-shell condition
\begin{equation}
*\Omega_{p+1}=-iW_{d-p-1}-g_p\delta{\Sigma_{p+1}}
\label{eq:4.7}
\end{equation}
\noindent{}and the dual action
\begin{equation}
S(V_{d-p-2})=\frac{1}{2}\int_{\cal{M}}W_{d-p-1}(V)\wedge
*W_{d-p-1}+g_{d-p-2}\int_{\Sigma_{d-p-2}} V_{d-p-2}
\label{eq:S-(d-p-2)-form}
\end{equation}
where
\begin{equation}
\int_{\Sigma_{d-p-2}}\cdot=-\frac{g_p}{g_{d-p-2}}
\int_{\cal{M}}d(*\delta(\Sigma_{p+1}))\cdot
\end{equation}
From (\ref{eq:4.2}) and (\ref{eq:4.7})  we obtain
the quantization condition
\begin{equation}
g_p g_{d-p-2}=2\pi{}n
\end{equation}

The quantum equivalence of the dual actions (\ref{eq:s-p-form})
and (\ref{eq:S-(d-p-2)-form}) follows once one integrates over all
corresponding higher order bundles. This is a generalization of
the equivalence proven in section (\ref{sec:u1}) for the
electromagnetic duality. The quantization of charges is directly
related to the different higher order bundles that may be
constructed over ${\cal{M}}$ and it arises naturally from the
global constraint (\ref{eq:global-omega}) needed for having a well
defined gerbe. The correspondence between closed integral p-forms
and bundles is in general not one-to-one, depending on the
topology of the base manifold, the redundancy being given by
$H^{p-1}({\cal{M}}, U(1))$~\cite{mario97}.

\section{ Global analysis of duality maps
in p-brane theories}
\label{sec:string}

We use in this section the  global arguments of the previous
sections to improve the p-brane $\Leftrightarrow$ $d$-brane
equivalence that has been proposed by~\cite{Town}
\cite{x6}\cite{x7}. The duality transformation has been used by
Townsend \cite{Town} to show the equivalence between the covariant
$D=11$ supermembrane action with one coordinate ($x^{11}$)
compactified on  $S^1$, and the fully $d=10$ Lorentz covariant
worldvolumen action for the $d=10$ IIA Dirichlet supermembrane.
The equivalence between the bosonic sectors was previously shown
by Schmidhuber \cite{x7} using the Born-Infeld type action found
by Leigh \cite{x6}. We will argue in a global way showing the
equivalence between both theories, even when nontrivial line
bundles are included in the construction of the D-brane action. We
discuss later on the equivalence of the bosonic sectors when the
coupling to background fields is included. Following the
Howe-Tucker formulation of the $d=11$ supermembrane~\cite{Town},
we consider a supermembrane sitting on a target manifold with one
coordinate compactified on $S^1$~\cite{Howe}, that is, we take
$x^{11}$ to be the angular coordinate $\varphi$ on $S^1$,
accordingly, we are interested in the following action
\begin{eqnarray}
S&=&-\frac{1}{2}\int_X d^{3}\xi
\sqrt{-\gamma}[\gamma^{ij}\pi_{i}^{m}\pi_{j}^{n}\eta_{mn}+\nonumber\\
&+&\gamma^{ij} (\partial_{i}\varphi
-i\bar{\theta}¥\Gamma_{11}\partial _{i}\theta)(\partial_{j}\varphi
-i\bar{\theta} \Gamma_{11}\partial _{j}\theta)-1]\nonumber\\
&-&\frac{1}{6}\int_X
d^{3}\xi\epsilon^{ijk}[b_{ijk}+3b_{ij}\partial_{k}\varphi]
\label{eq:membrane-action}
\end{eqnarray}
where $\eta$ is the Minkoswski metric on a $10$-dimensional
spacetime, and
\begin{eqnarray} \pi^{m} & = &
dx^{m}-i\bar{\theta}\Gamma^{m}d\theta\\
\epsilon^{ijk}b_{ijk}&=&3\epsilon^{ijk}\{
i\bar{\theta}\Gamma_{mn}\partial_{i}\theta[\pi_{i}^{m}\pi_{j}^{n}+
i\pi_{i}^{m}(\bar{\theta}\Gamma^{n}\partial_{j}\theta)-\nonumber\\
&-&\frac{1}{3}(\bar{\theta}\Gamma^{m}\partial_{i}\theta)
(\bar{\theta}\Gamma^{n}\partial_{j}\theta)]+\nonumber\\
&+&(\bar{\theta}\Gamma_{11}\Gamma_{m}\partial_{i}\theta)(\bar{\theta}
\Gamma_{11}\partial_{j}\theta)
(\partial_{k}x^{m}-\frac{2i}{3}\bar{\theta}\Gamma^{m}\partial_{k} \theta)\}\\
\epsilon^{ijk}b_{ij}&=&-2i\epsilon^{ijk}\bar{\theta}\Gamma_{m}\Gamma_{11}
\partial_{i}\theta(\partial_{j}x^{m}-\frac{i}{2}\bar{\theta}\Gamma^{m}\partial_{j}\theta).
\end{eqnarray}

In order to discuss duality we will follow the approach we have
introduced in the preceding sections. We begin by letting
$\Sigma_1$ be a basis of homology on the three dimensional
worldsheet manifold ($X$), $L_1$ be a $1$-form satisfying the
constraints
\begin{eqnarray}
dL_{1}&=&0\\
\oint_{\Sigma_{1}}L_1 & = & 2\pi{}n,\qquad{}n\in\entero
\end{eqnarray}
then, as we have learnt, $L_1$ defines a class of uniform maps
 $X\rightarrow{}S^1$ ($-1$-gerbe) via
\begin{eqnarray}
g&=&\exp{i\varphi},\quad\mbox{and}\\
L_1&=&-ig^{-1}dg=d\varphi.
\end{eqnarray}
The converse being also valid.

The next step in the construction of the duality map consists then
in attaining an equivalent formulation to the action
(\ref{eq:membrane-action}) in terms of the global $1$-form $L_1$.
In order to achieve this goal we must notice that the Lagrange
formulation of the constraints on $L_1$ may be obtained in terms
of a connection $1$-form over the space of all non trivial line
bundles, i.e. the global constraints can be included in the action
by introducing an auxiliary line bundle with connection and
coupling its curvature with $L_1$ (which we regard as globally
defined but unconstrained) in the following fashion
\begin{equation}
\int_{X}dF(A)\wedge{}L_1
\end{equation}

According to this, we begin the discussion of duality by
introducing the following master action (here $L_{1i}$ stands for
the i-component of $L_1$)
\begin{eqnarray}
S_{Master}&=&-\frac{1}{2}\int_{X}d^{3}\xi\sqrt{-\gamma}
[\gamma^{ij}\pi_{i}^{m}\pi_{j}^{n}\eta_{mn}+\nonumber\\
&+&\gamma^{ij}(L_{1i}-i\bar{\theta}\Gamma_{11}\partial _{i}\theta)
(L_{1j} -i\bar{\theta}\Gamma_{11}\partial_{j}\theta)-1]-\nonumber\\
&-&\frac{1}{6}\int_{X}d^{3}\xi\epsilon^{ijk}
[b_{ijk}+3b_{ij}L_{1k}]+\int_{X}F(A)L_{1}
\label{eq:master}
\end{eqnarray}
Where $A$ is a connection on a line bundle over $X$ and where the
path integral must sum over all connections and all line bundles.

Functional integration of the exponential of the master action on
$A$ yields the following factor on the measure (recall the
arguments in section 2)
\begin{equation}
\delta(dL_{1})\delta(\oint_{\Sigma_{1}}L_1 - 2\pi n)
\end{equation}
We now use the fact that
\begin{equation}
\delta(dL_{1})\delta(\oint_{\Sigma_{1}}L_1-2\pi n)=
\int [d\varphi]\frac{\delta(L_1-d\varphi )}{\det{}d}
\label{eq:5.9}
\end{equation}
where $\varphi$ defines a map from $X\rightarrow{}S^1$, this shows
that $d\varphi$ satisfies the constraint on the periods. At this
point we notice that the functional integral appearing in
(\ref{eq:5.9}) is over all maps from $X\rightarrow S^1$, in other
words, it is not an integration over a cohomology class defined by
an element of $H^1(X)$. In distinction to section 2, the zero
modes in this case, are constants. We may hence directly integrate
the path integral associated to $S_1$ on $L_1$ and replace $L_1$
by $d\varphi$ a choice that leads us to the covariant $d=11$
supermembrane action.

On the other hand, we might have functionally integrated over
$L_1$ in to arrive to the functional integral of the action
\begin{eqnarray}
S_{2}&=&-\frac{1}{2}\int_{X}d^{3}\xi
\sqrt{-\gamma}[\gamma^{ij}\pi_{i}^{m}\pi_{j}^{n}
\eta_{mn}-\gamma^{ij}f_i(A) f_j(A)-1]\nonumber\\
&-&\frac{1}{6}\int_{X}d^{3}\xi\epsilon^{ijk}b_{ijk}+
\int_{X}d^{3}\xi\gamma^{ij}f_i(A)i\bar{\theta}
\Gamma_{11}\partial_{l}\theta
\label{eq:5.10}
\end{eqnarray}
Where
\begin{equation}
f_i(A)\equiv\epsilon_{imn}(F^{mn}(A)-\frac{1}{2}b^{mn}).
\end{equation}

The functional integral in $A$ must now be performed over all line
bundles over $X$. The result (\ref{eq:5.10}) was obtained by
Townsend in {\cite{Town}}, for the case of a trivial line bundle.
The equivalence between (\ref{eq:5.10}), the fully $d=10$ Lorentz
covariant worldvolume action for the $d=10$ IIA Dirichlet
supermembrane, and the $d=11$ covariant supermembrane action
(\ref{eq:membrane-action}) has then been established.  In the
functional integral for (\ref{eq:membrane-action}), integration
over all maps between $X\rightarrow{}S^1$ must be performed while
in the functional integral for (\ref{eq:5.10}) the integration
must be over the space of all connection $1$-forms on all line
bundles (modulo gauge transformations).

The global aspects of (\ref{eq:5.10}) are even more interesting
when the coupling of the formulations to background fields is
considered. In the $d=10$ membrane action obtained by dimensional
reduction of the $d=11$ membrane theory the local $2$-form $B$ of
the NS-NS sector, couples to the current
$\epsilon^{ijk}\partial_k\varphi$.  The coupling is a topological
one. Assuming we are in the euclidean worldvolume formulation of
the theory, the coupling admits sources $B$ which are locally
2-forms but globally associated to nontrivial higher order gerbes.
The reformulation of the action in terms of $1$-forms $L_{1}$ and
constraints $dL_1=0$ and $\int{}L_1\sim{}n$ is close to what was
done to obtain $S_2$ from the master action (\ref{eq:master}), but
substituting $f_{i}(A)$ by
\begin{equation}
\hat{f}_i(A)\equiv \epsilon_{imn}(F^{mn}(A)+B^{mn})
\end{equation}
where only the bosonic sector is considered. There is an
interesting change in procedure, however, arising from the
nontrivial transitions of $B$. The result is that $F^{mn}(A)$ must
have also nontrivial transitions that compensate the ones of $B$.
We have in the intersection of two opens $U^\prime \cap U \neq
\phi$ where a nontrivial transition takes place
\begin{eqnarray}
B^{\prime}&=&B+d\eta\nonumber\\
F^{\prime}&=&F-d\eta
\end{eqnarray}
which imply
\begin{equation}
A^\prime= A-\eta.
\end{equation}

This new transition for the connection 1-form A arises naturally
in the topological field actions introduced in {\cite{Martin}} to
describe a gauge principle from which the Witten-Donaldson and
Seiberg-Witten invariants may be obtained as correlation functions
of the corresponding BRST invariant effective action. The most
appropriate theory, however, where the nontrivial p-form
connections are expected to have  relevant non perturbative
effects is the $d=11$ 5-brane. It has been conjectured
{\cite{Town}} that the $d=11$ 5-brane action is given by
\begin{equation}
S=-\frac{1}{2}\int_{X} d^{6}\xi
\sqrt{-\gamma}[\gamma^{ij}\partial_{i}x^{M}
\partial_{j}x^{N}\eta_{MN}+\frac{1}{2}\gamma^{il}\gamma^{jm}\gamma^{kn}F_{ijk}F_{lmn}-4]
\end{equation}
where $F=dA$ is the self dual $3$-form field strength of a local
$2$-form potential $A$, consequently, the context for this
discussion is that of $1$-gerbes. There is a very rich geometrical
structure associated to this action with non perturbative effects
related to the non trivial gerbes. The $d=11$ $5$-brane has been
also interpreted {\cite{Town}} as a Dirichlet-brane of an open
supermembrane, with boundary in the $5$-brane worldvolume
described by a new six-dimensional superstring theory previously
conjectured by \cite{Witten3}.  We expect that these intrinsic
non-perturbative effects should be realized naturally over
non-trivial higher order gerbes.

\section{Conclusions}
\setcounter{equation}{0}

In this article we have introduced the notion of gerbes which are
the natural setting to discuss $p$-forms with local transitions.
We have also shown that gerbes allow  a global extension of
duality transformations in quantum field theory.

Our approach to duality incorporates globally constrained forms
which give raise to the dual gerbes. The constraints can be easily
incorporated into suitable master actions, care should be taken
since the incorporation of the constraints has a subtlety
associated with integration on boundaries i.e. stokes theorem. The
constraints include relevant physical parameters such as coupling
constants associated to the interaction of the $p$-forms to the
underlying $p$-branes, or the radius of compactification of the
superstring or supermembrane. The dependence becomes relevant in
proving quantum equivalence between dual string and membrane
theories.

Finally, we presented an improvement, of the equivalence between
the covariant $d=11$ supermembrane action with one coordinate
compactified on $S^1$ and the fully $d=10$ Lorentz covariant
worldvolume action for the $d=10$ IIA Dirichlet supermembrane
which even includes some global aspects.

\vskip.5cm

\noindent{\large\bf{Acknowledgements}}

M.C. would like to acknowledge the hospitality of the Center of
Theoretical Physiscs (C.T.P.) , Laboratory for Nuclear Science and
Department of Physics of the Massachusetts Institute of
Technology.

The financial support for this project came from several sources,
we received partial funding from project $G-11$ form the
{\it{Decanato de Investigaciones de la Universidad Sim\'on
Bol\'{\i}var}}. The visit of M.I.C. to C.T.P. was made possible by
a sabbatical grant form Universidad Sim\'on Bol\'{\i}var, and also
by partial support from funds provided by the U.S. Department of
Energy (D.O.E.) under cooperative research agreement
DF-FCO2-94ER40810.

\end{document}